\documentclass[12pt]{iopart}
\input{psfig.sty}

\def\apj{{ApJ}}

\newcommand{\apjs}{{Astrophys.~J.~Supp.}}

\newcommand{\mnras}{{Mon.~Not.~R.~Astron.~Soc.}}
\newcommand{\prd}{{Phys. Rev. D}}
\newcommand{\prl}{{Phys. Rev. Lett.}}
\newcommand{\isw}{{ISW}}
\newcommand{\Ylmn}{{Y_{lm}}}

\newcommand{\deld}{{\delta^D}}
\newcommand{\rad}{{r}}
\newcommand{\bn}{\hat{\bf n}}
\newcommand{\veck}{{\bf k}}
\newcommand{\be}{\begin{equation}}
\newcommand{\ee}{\end{equation}}
\newcommand{\bea}{\begin{eqnarray}}
\newcommand{\eea}{\end{eqnarray}}

\begin{document}
%opening
\title[Integrated Sachs-Wolfe Effect in CMB Polarization]{Searching For Integrated Sachs-Wolfe Effect Beyond Temperature Anisotropies: CMB E-mode 
Polarization-Galaxy Cross Correlation}
\author{Asantha Cooray}

\address{Dept. of Physics \& Astronomy
 University of California,
 Irvine, CA 92697-4575}

\author{Alessandro Melchiorri}
\address{Dipartimento di Fisica ``G. Marconi'' and INFN, sezione
  di Roma, Universita' di Roma ``La Sapienza'', Ple Aldo Moro 5,
  00185, Roma, Italy.}

\begin{abstract}
The cross-correlation between cosmic microwave background (CMB) temperature anisotropies and the
large scale structure (LSS) traced by the galaxy distribution, or sources at different wavelengths,
is now well known. This correlation results from the integrated Sachs-Wolfe (ISW) effect in CMB anisotropies
generated at late times due to the dark energy component of the Universe.
In a reionized universe, electron scattering at low redshifts leads to a large-scale polarization contribution.
In addition to the primordial quadrupole, involving anisotropies at the last scattering surface,
the ISW quadrupole rescatters and contributes to the large angular scale polarization signal.
Thus, in principle, the low multipole polarization bump in the E-mode should be correlated with the galaxy distribution.
Unlike CMB temperature---LSS correlation that peaks for tracers at low redshifts,
due to the increasing visibility function, while the fractional ISW quadrupole is decreasing, with increasing redshift,
the polarization---galaxy correlation peaks mostly at redshifts between 1 and 3. 
Under certain conditions, mostly involving a low optical depth to reionization
if the Universe reionized at a redshift around 6, the cross polarization---source signal is marginally detectable if
all-sky maps of the large scale structure at redshifts between 1 and 3 is available.
If the Universe reionized at a redshift higher than 10, it is unlikely that this correlation will be detectable
even with no instrumental noise all-sky maps.
While our estimates do not guarantee a detection, unknown physics related to the dark energy
uncertain issues related to the large angular scale CMB and polarization anisotropies, and uncertainties related to
the reionization history of the Universe may motivate attempts to measure this correlation using upcoming CMB polarization E-mode maps
and large-scale structure maps of the high-redshift universe with Large Synoptic Survey Telescope and 
quasar catalogs, among others.
\end{abstract}

\maketitle

\section{Introduction}
\label{sec:intro}

At large angular scales, in addition to primordial  cosmic microwave background (CMB) anisotropies generated at the last scattering surface,
the integrated Sachs-Wolfe (ISW; Sachs \& Wolfe 1967)  effect also provides an additional contribution at late times.
The temperature fluctuations in the ISW effect results from the 
differential redshift effect from photons climbing in and out of time
evolving potential perturbations from last scattering surface to
present day\footnote{The ISW effect generates anisotropies at two epochs: at early times during the transition from radiation domination
to matter domination and at late times during the transition from matter domination to dark energy domination.
In this paper, we will refer to ISW as the late time contribution, while the early time effect is considered as part of
the primordial quadrupole and primordial anisotropies.}.
In currently popular cold dark matter cosmologies with a 
cosmological constant, significant contributions to temperature anisotropies arise
at redshifts less than 1 and on and above the scale of the horizon at the time of decay.
It is then expected that the ISW effect is correlated with the
large scale structure (e.g., Crittenden \& Turok 1996). Since the ISW contribution is sensitive to how one models
cosmology at late-times, such as the presence of a dark energy component or not,
the correlation between the ISW effect and the tracers of the
large scale structure has been suggested as a probe of the dark energy properties (Corasaniti et al. 2003;
Hu \& Scranton 2004; Garriga et al. 2004; Pogosian 2004; Pogosian et al. 2005).

Several attempts have now been made to detect the correlation between
the ISW effect, as traced by CMB maps of the Wilkinson Microwave Anisotropy Probe (WMAP; Bennett et al. 2003),
 and sources at optical (Scranton et al. 2003; Padmanabhan et al. 2005; 
Afshordi et al. 2004; Fosalba \& Gaztanaga 2004; Fosalba et al. 2003),  X-ray, and radio (Nolta et al. 2004; Boughn \& Crittenden 2004)
 wavelengths. Detections are at the level of 2 to 3 $\sigma$ depending on
what sources are used and what their redshift distributions are, among other parameters related to the tracer field
and whether the statistic is in Fourier space or real space.
These detections have allowed certain constraints on dark energy
parameters (e.g., Corasaniti et al. 2005) and additional applications such as a way to slightly improve
the tensor-to-scalar ratio (Cooray et al. 2005).  Looking for additional signatures of the ISW effect 
(e.g., Cooray \& Baumann 2003) is useful to further improve our understanding of dark energy 

While temperature anisotropies have been used so far to look for the ISW effect, it is
possible that CMB polarization anisotropies also contain signature of time-evolving potential
perturbations at low redshifts compared to the redshift of last scattering.
Linear polarization of the CMB is generated essentially
through rescattering of the temperature quadrupole (Rees 1968).  At the last scattering surface, this
leads to the primordial polarization contribution (e.g., Hu \& White 1997),  while at lower  redshifts, the 
temperature  quadrupole gets scattered again when the Universe reionizes.
This late-time reionization contribution generates a large angular scale polarization corresponding to the 
angular scale of the horizon  at the surface of reionization (Zaldarriaga 1997). 

This apparent excess of polarization at large angular scales, as expected in a reionized Universe at late-times,
has now been detected with WMAP using the temperature-polarization cross power spectrum (Kogut et al. 2003).
Models of this signal including other cosmological parameters indicate that the optical depth to reionization is 
around $\tau \sim 0.17$ such that the redshift of reionization is around 20. However, there are large degeneracies between the optical
depth and other parameters as well as large uncertainties resulting from the low signal-to-noise level of this detection.
Moreover, foregrounds in the CMB polarization maps are yet to be characterized properly. The expected improvements of WMAP data 
over the next few years as well as the 
planned Planck experiment will significantly improve the signal-to-noise level of CMB polarization maps in the near future.

Since the temperature anisotropy quadrupole is rescattered to generate a polarization signal, 
the large angular scale polarization signal, in principle, should contain the quadrupole generated by the
ISW effect in addition to the dominant
quadrupole associated with primordial fluctuations at the last scattering surface. Again, in principle,
large angular scale polarization  should be correlated with tracer fields of the large-scale structure
 at which the polarization signal related ISW quadrupole is generated. 
Since the scattering probability increases with increasing redshift while
the ISW contribution is decreasing, unlike temperature anisotropies from the ISW that peak at redshifts $< 1$ (Cooray 2002b;
Afshordi 2004),
ISW polarization contribution is expected to peak at a slightly higher redshift. Thus, tracer fields at redshifts $\sim$ 1 to 3 are more
likely to be suitable for the cross-correlation. 

However, a reliable detection of this cross-correlation must overcome several challenges.
For example, a large fraction of the polarization signal is generated by
scattering of the primordial quadrupole involving potential perturbations at the last scattering surface
associated with the Sachs-Wolfe (SW) effect. To detect the ISW effect in the form of a cross-correlation with sources at the ideal
redshift range, SW contribution to polarization should be smaller such that the cumulative polarization is not dominated by
primordial quadrupole but rather by the late-time ISW.  In the case of temperature anisotropies, in certain
cosmological models, the SW and ISW contributions to large angular scales CMB anisotropies are at the same order 
of magnitude. In the case of polarization, the ISW contribution to the large angular scale signal
is at the level of $\sim$ 0.01 $\mu$K or below, depending on the background cosmology. 
In comparison, if the optical depth to reionization is $\tau > 0.08$, the scattering of the primordial quadrupole
lead to a polarization signal higher than 0.1 $\mu$K. Thus, the detection of ISW contribution in polarization is challenging
since the primordial polarization results in a large noise source within which correlations with tracer fields are measured.

The detection may be facilitated under certain conditions. For example, if
 the overall optical depth to reionization is small, say if reionization was at a redshift around 6
as suggested by some of the optical observations (e.g., Fan et al. 2001; White et al. 2003),
then the ratio of ISW-to-SW signal in the polarization would be higher, say when compared to a
universe reionized at a redshift greater than 10. Assuming standard
LCDM cosmology with the dark energy described by a cosmological constant, based on the numerical calculations presented here,
we find that it is possible to make a marginal detection of the ISW signal in polarization in this case.
There may be other possibilities that increase the signal-to-noise ratio of the detection. 
If dark energy, which is marginally understood with current data, could
behave in a way that it's behavior leads to a slightly higher ISW contribution at higher redshifts than expected in models of
a cosmological constant. While we have not investigated such a possibility here,
various models that are developed in the literature could potentially be tested. 
The third possibility that may improve a detection is changes to the primordial quadrupole than what is expected in
standard cosmological models. For example, 
large angular scale anomalies in CMB temperature anisotropy data, both from COBE and WMAP,
suggests an apparent lack of primordial power at largest angular scales (e.g., Spergel et al. 2003; Efstathiou 2003;
Copi et al. 2005).
Such a decrease in power is expected to make slight modifications to the polarization signal and to result in 
a lower polarization signal at large angular scales than assumed with no modifications to the primordial spectrum 
(e.g., Cline et al. 2003). In fact, CMB polarization towards galaxy clusters may
 provide a mechanism to look for 
the cut-off in primordial power spectrum (e.g., Baumann \& Cooray 2003; Seto \& Pierpaoli 2005).
If there is a cut-off in the spectrum at small scales, ISW signal remain unaffected and 
the ratio of ISW-to-SW contribution to polarization may be higher when compared to a model where large-scale power is not affected. Thus, we find that there are good motivations to look for the ISW signature in 
polarization anisotropies through cross-correlations with sources in addition to same studies with temperature anisotropy maps alone.

This paper is organized as following: in the next Section, we briefly summarize the ISW signal in
CMB polarization at large angular scales and the cross-correlation between a tracer field of the large scale structure such as galaxies.
We also estimate signal-to-noise levels for a detection of this correlation
and show that the overall detectability is marginal, with a signal-to-noise ratio of about 2 with all-sky maps of the tracer field,
if the universe reionized at a redshift around 6. This ratio could potentially improve in certain models of dark energy, especially ones where
a significant derivative exists for the dark energy equation of state at redshifts between 1 to 3.
Moreover, the detectability may also improve under conditions where the primordial contribution is smaller, such as in models where the large-scale power has a cut-off; Though numerical calculations with a cut-off show that the overall increase in the signal-to-noise ratio is
insignificant since it only boosts the correlation level of lowest order multipoles.
Finally, the signal-to-noise ratio we quote is for the detection in terms of the angular cross power spectrum between
galaxies and the polarization E-mode, techniques could be improved to  detect this cross signal in pixel space or
using other data analysis techniques where the primordial contribution is accounted for and filtered out.
We conclude with a summary of our results.
For illustration purposes, we assume a cosmological model with $\Omega_m=0.25$, $\Omega_\Lambda=0.75$, $h=0.72$, and
a normalization to the power spectrum of $\sigma_8=1.0$.

\section{Calculational Method}

To describe the existence of a correlation, we first summarize the ISW contribution to CMB anisotropies.
The ISW effect (Sachs \& Wolfe 1967) results from the late time decay of gravitational potential fluctuations. The
resulting temperature fluctuations in the CMB can be written as
\begin{equation}
T^\isw(\bn) = -2 \int_0^{\rad_s} d\rad \dot{\Phi}(\rad,\bn \rad) \, ,
\end{equation}
where the overdot represent the derivative with respect to conformal
distance (or equivalently look-back conformal time) and $\Phi(\rad,\bn \rad))$ is the potential perturbations in the large-scale structure.
Here, $\rad_s$ is the radial distance to the last scattering surface.

Writing multipole moments of the temperature fluctuation field
$T(\hat{\bf n})$,
\begin{equation}
a_{lm} = \int d\bn T(\bn) \Ylmn {}^*(\bn)\,,
\end{equation}
we can formulate the angular power spectrum as
\begin{eqnarray}
\langle a_{l_1m_1}^* a_{l_2m_2}\rangle = \deld_{l_1 l_2} \deld_{m_1 m_2}
        C_{l_1}\,.
\end{eqnarray}
For the ISW effect, multipole moments are
\begin{eqnarray}
a^{\rm ISW}_{lm} &=&i^l \int \frac{d^3\veck}{2 \pi^2}
\int d\rad   \delta(\veck) I_l(k)  \Ylmn(\hat{\veck}) \, ,
\nonumber\\
\label{eqn:moments}
\end{eqnarray}
with $I_l(k) = \int d\rad W^\isw(k,\rad) j_l(k\rad)$ with the window
function for the ISW effect
\begin{equation}
W^\isw(k,\rad) = -3 \Omega_m \frac{H_0^2}{k^2} \frac{d}{d\rad}\left(\frac{G}{a}\right) \, .
\end{equation}
Here, we have made use of the Poisson equation to related potential fluctuations to the density field and have assumed
linear perturbations with the growth factor given by $G(\rad)$ (see, Cooray 2002b for details). 
We ignore non-linear corrections to the ISW effect since
they only lead to anisotropies at small scales and do not modify the quadrupole (Cooray 2002a). The function
$d/dr(G/a)$ is strongly sensitive to dark energy density and related parameters, 
while zero in an Einstein-de Sitter universe with $\Omega_m=1$.
As is well-known the overall ISW contribution provides a sensitive probe of dark energy. In fact, as discussed in
Cooray et al. (2004), if the ISW contribution can be established as a function redshift, it provides a more powerful probe of
dark energy than estimates involving luminosity distances and other cosmological probes (assuming they are also
measured to the same accuracy as a function of redshift).

\begin{figure*}[!t]
\centerline{\psfig{file=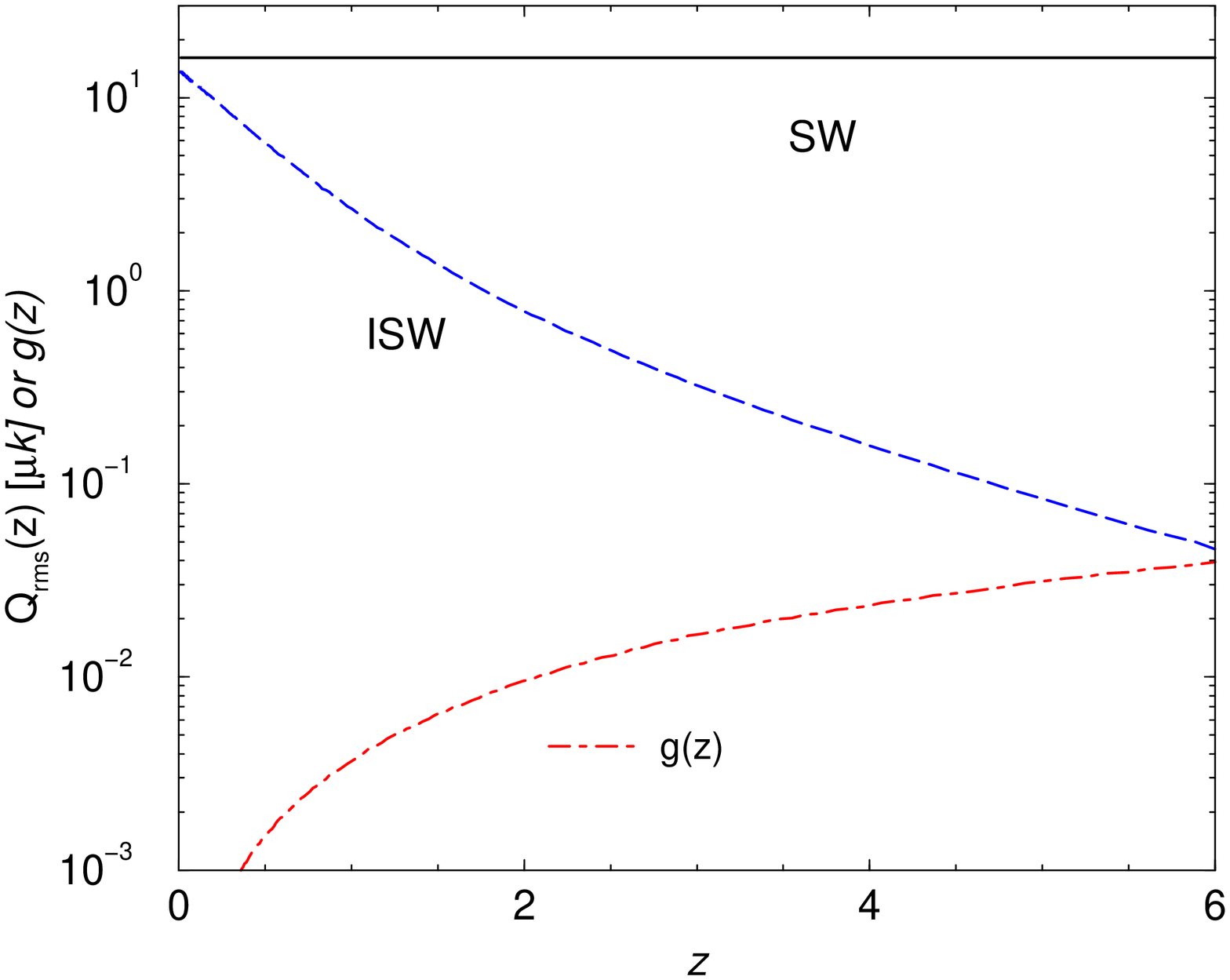,width=3.6in,angle=-90}
\psfig{file=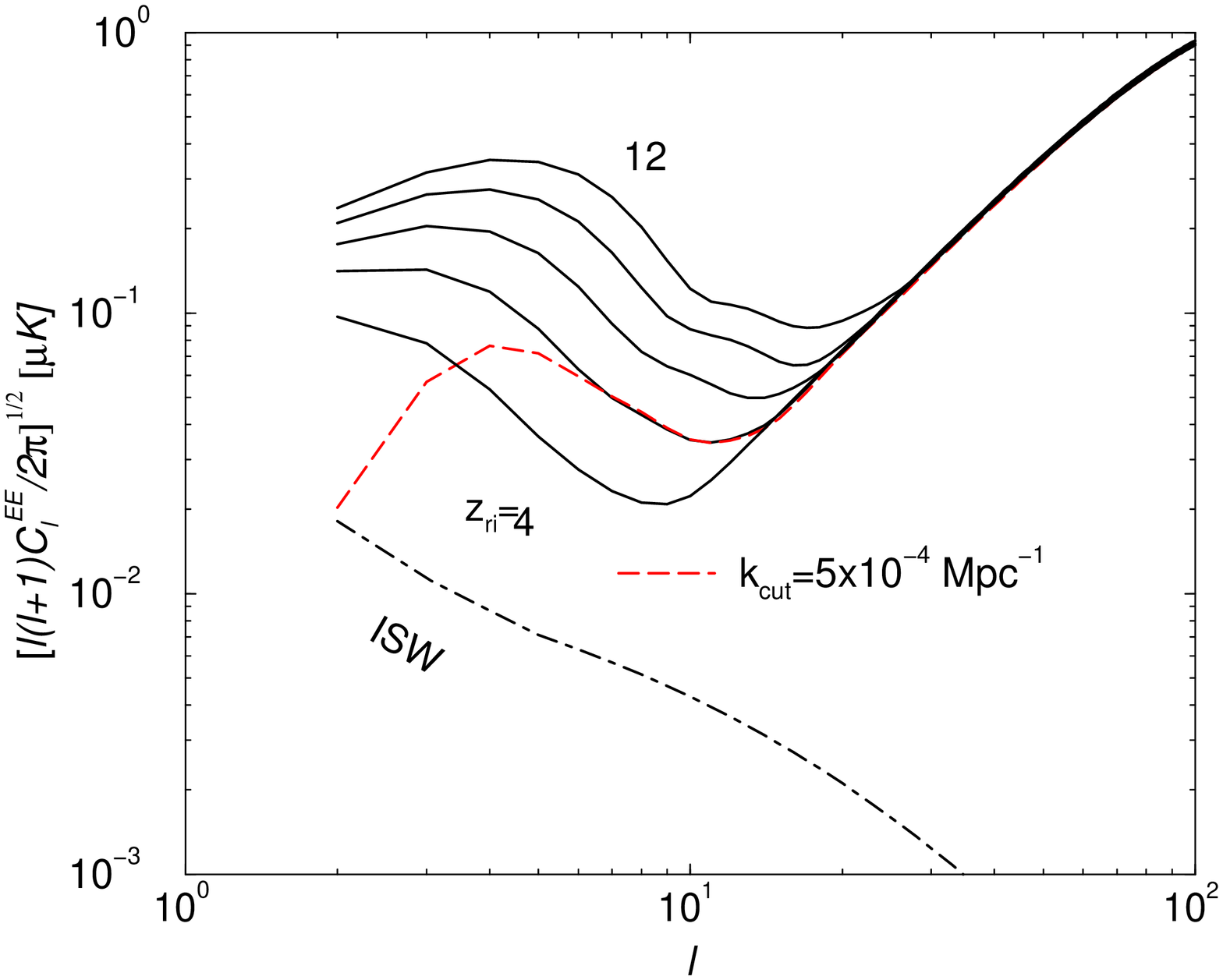,width=3.6in,angle=-90}}
\caption{{\it Left panel:} The rms quadrupole as a function of redshift. The top horizontal line is the primordial
quadrupole related to SW effect while the long-dashed line shows the ISW contribution that peak as the redshift is decreased.
For comparison on the redshift dependence, with dot-dashed line
we also show $g(z)$, the visibility function that determine the scattering probability.
This function increases as the redshift is increased, but sharply drops to zero beyond the redshift at which the Universe reionized.
Most of the scattering happens at the reionization redshift and the resulting polarization signal is dominated by scattering
of the primordial quadrupole. {\it Right panel:} The polarization anisotropy power spectrum, in the E-mode,
as a function of the multipole and as a function of the reionization redshift, from 4 to 12 at steps of 2.
Note that the ISW contribution is significantly smaller than the contribution from
SW quadrupole scattering. In fact, the ratio of ISW-to-SW contributions to polarization decreases as the reionization
redshift is increased. A favorable detection of the ISW effect in polarization, through a cross-correlation,
generally require a low optical depth or a reionization redshift $\sim$ 6. For reference, we also show the case
where the primordial power spectrum has a cut-off at wavenumbers below $5 \times 10^{-4}$ Mpc$^{-1}$, as required to
explain the low power of the temperature anisotropy quadrupole and the octupole (Kesden et al. 2003).}
\end{figure*}

\begin{figure*}[!t]
\centerline{\psfig{file=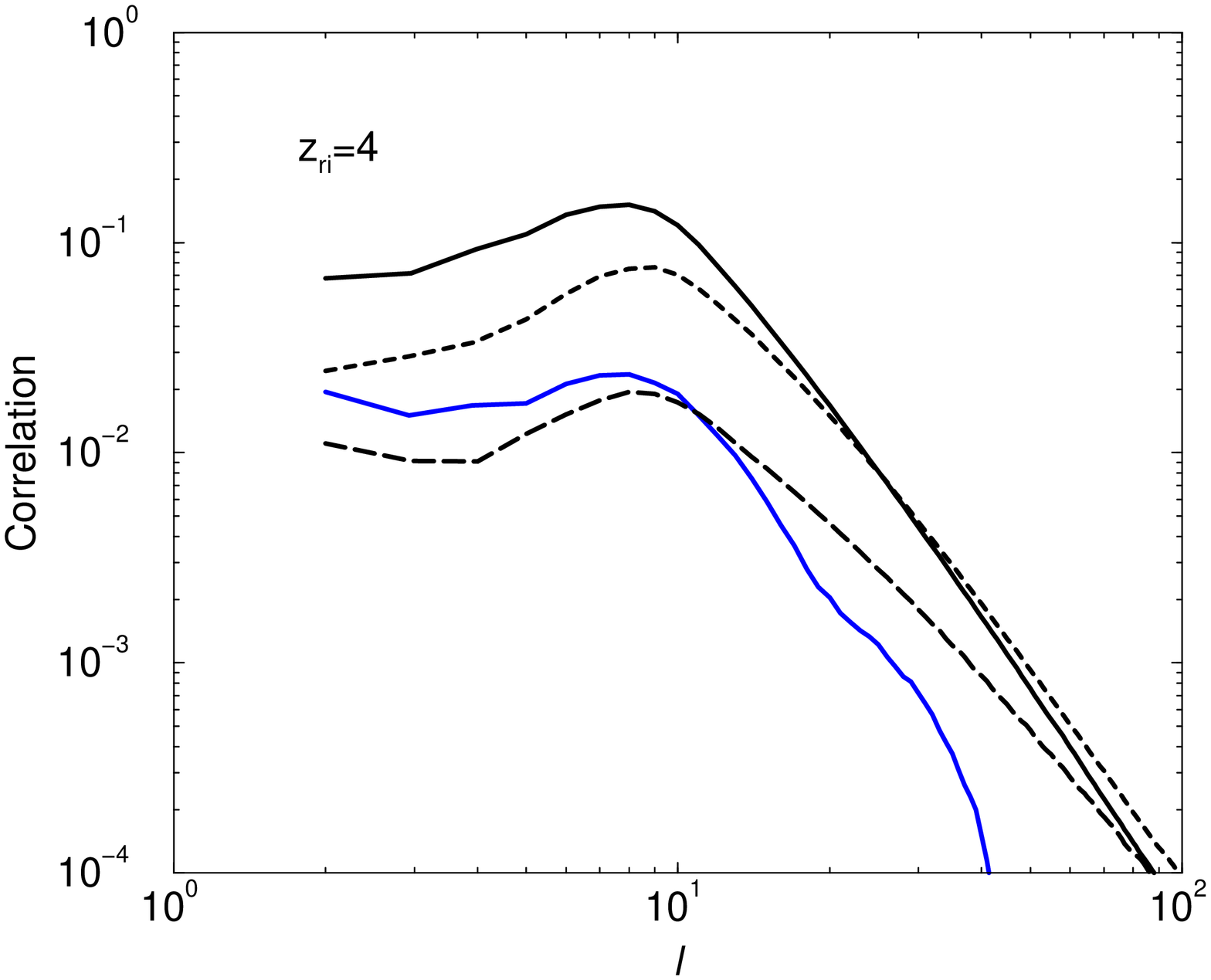,width=3.6in,angle=-90}
\psfig{file=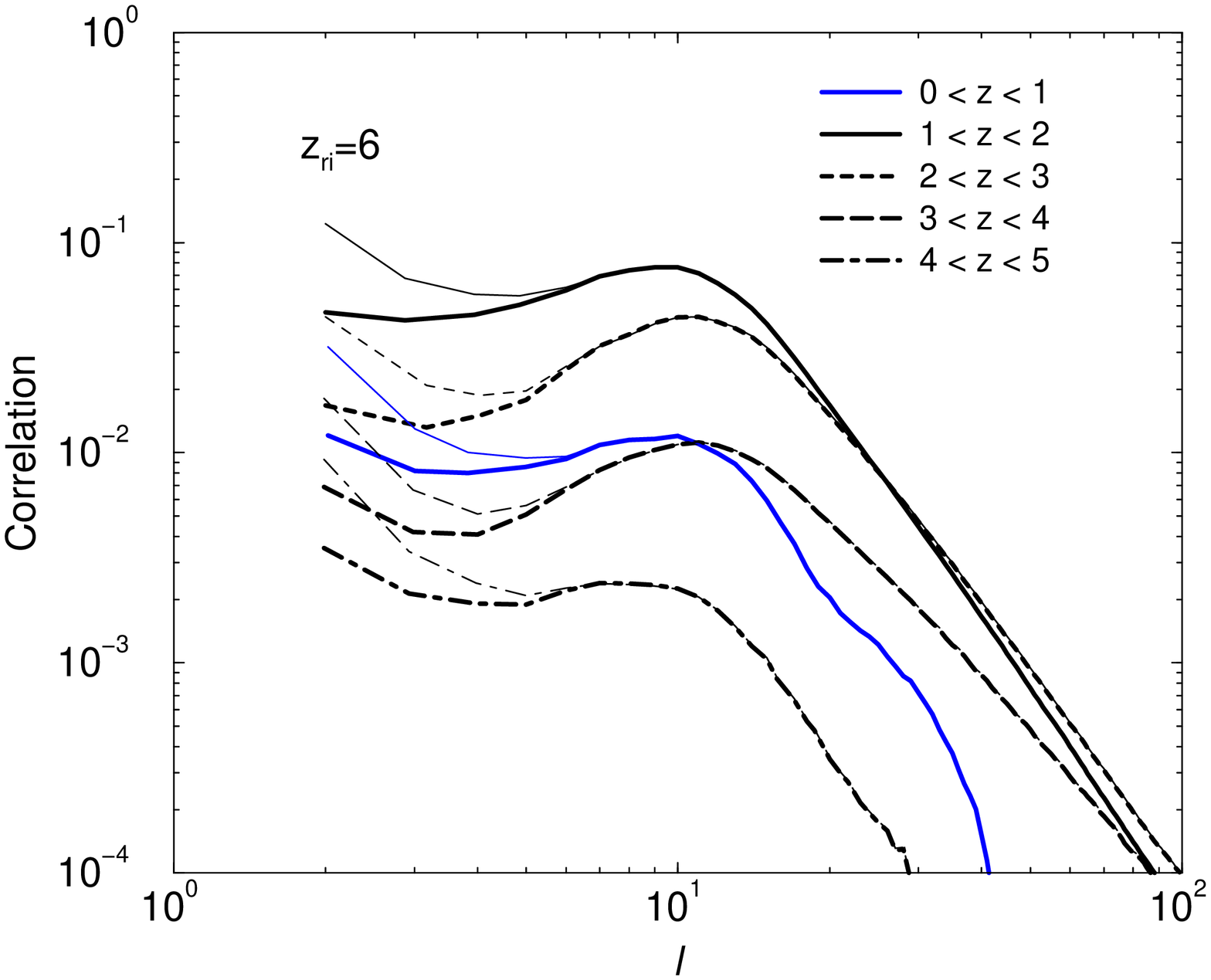,width=3.6in,angle=-90}}
\centerline{\psfig{file=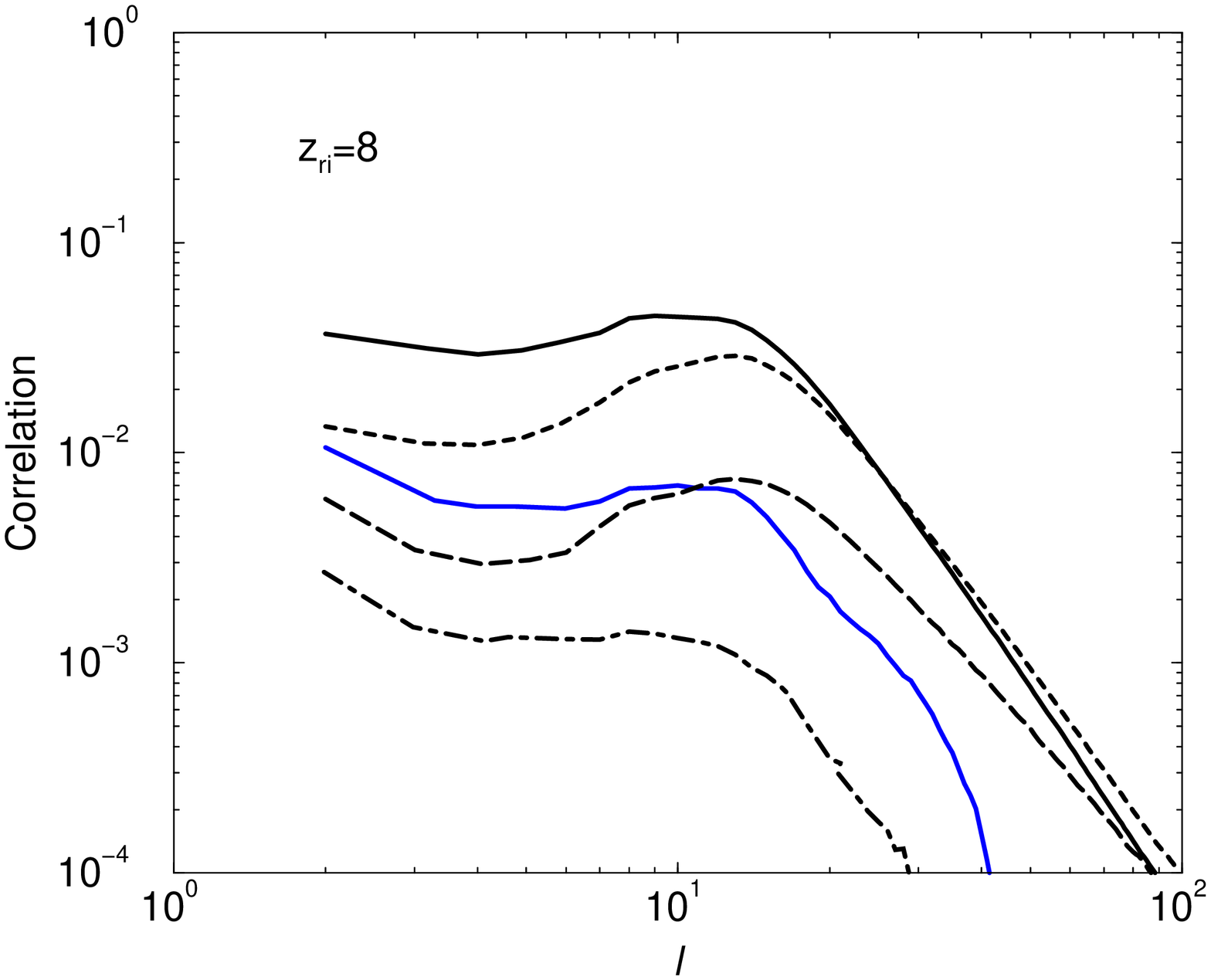,width=3.6in,angle=-90}
\psfig{file=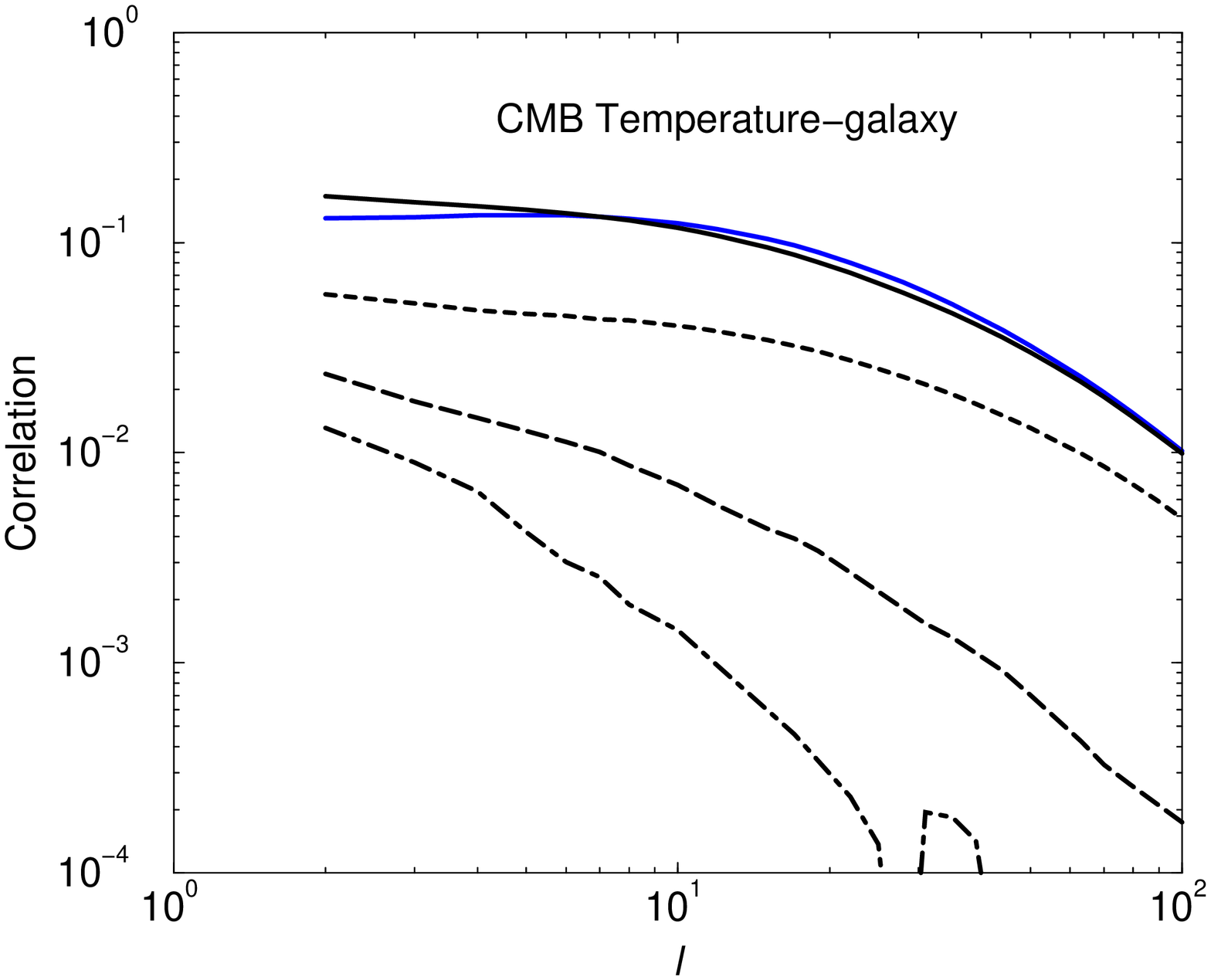,width=3.6in,angle=-90}}
\caption{The correlation coefficient $r$  between E-mode polarization maps  and a  tracer field, as a function of the redshift bin
of the source field. From top-left to bottom-left we show correlation coefficients assuming
a reionization redshift of 4, 6, and 8, respectively. The bottom-right panel shows the correlation  coefficient
between CMB temperature anisotropy maps and tracer fields again divided to redshift bins
as indicated on the top right panel. The correlations peak at a multipole $\sim 10$ and
decreases sharply thereafter due to the increase in $C_l^{\rm EE}$ when $l > 10$ (see. Figure~1 right panel).
In the case of polarization, correlation coefficients are higher for tracer fields at redshifts between 1 and 3, relative to coefficients for sources between redshifts of 0 and 1. This can be compared to
temperature-source correlation which peaks for sources between redshifts of 0 and 2.
In the top right panel, thin lines show the correlation coefficient with a cut-off in the power spectrum (see, Figure~1 right panel).
}
\end{figure*}

\begin{figure*}[!t]
\centerline{\psfig{file=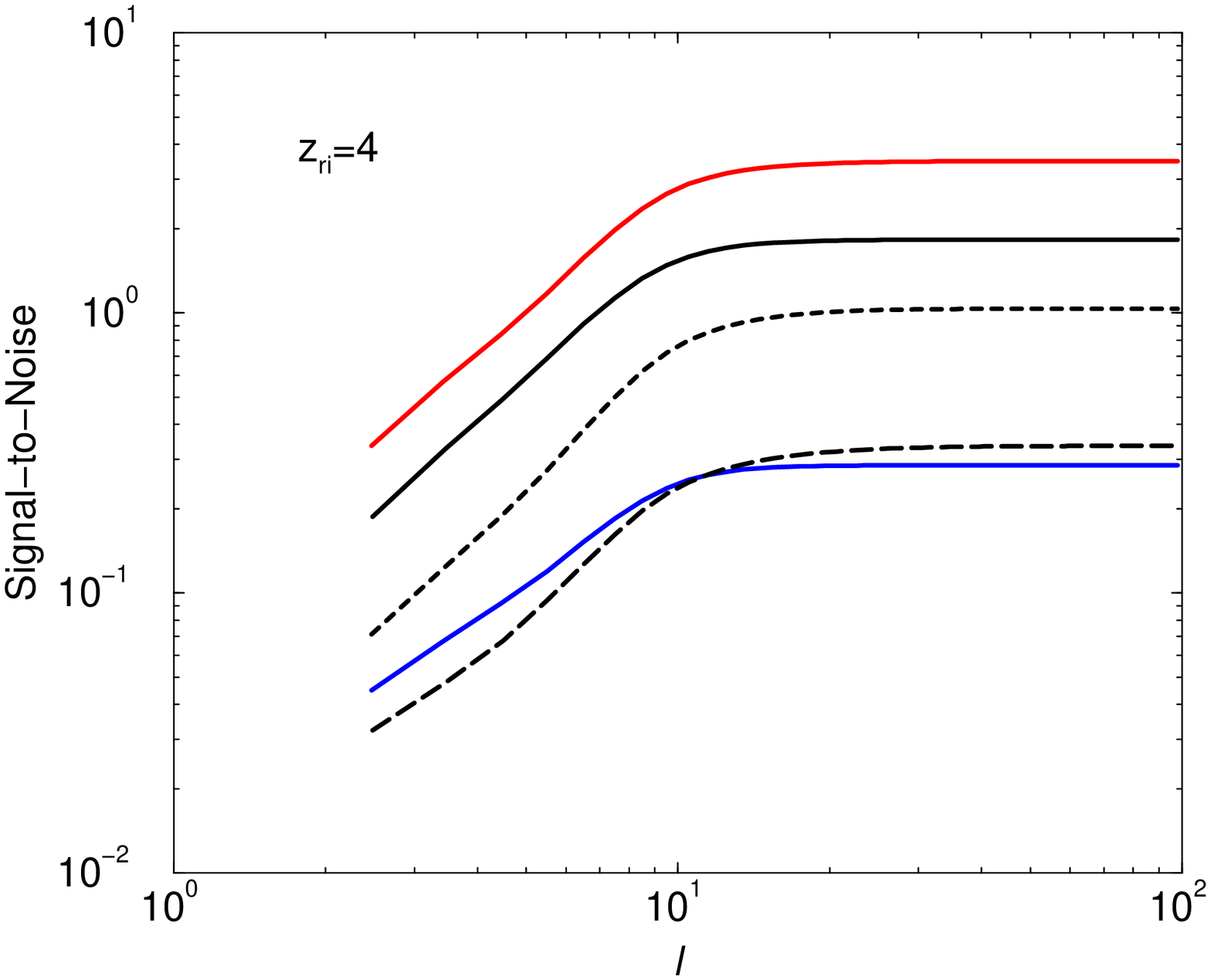,width=3.6in,angle=-90}
\psfig{file=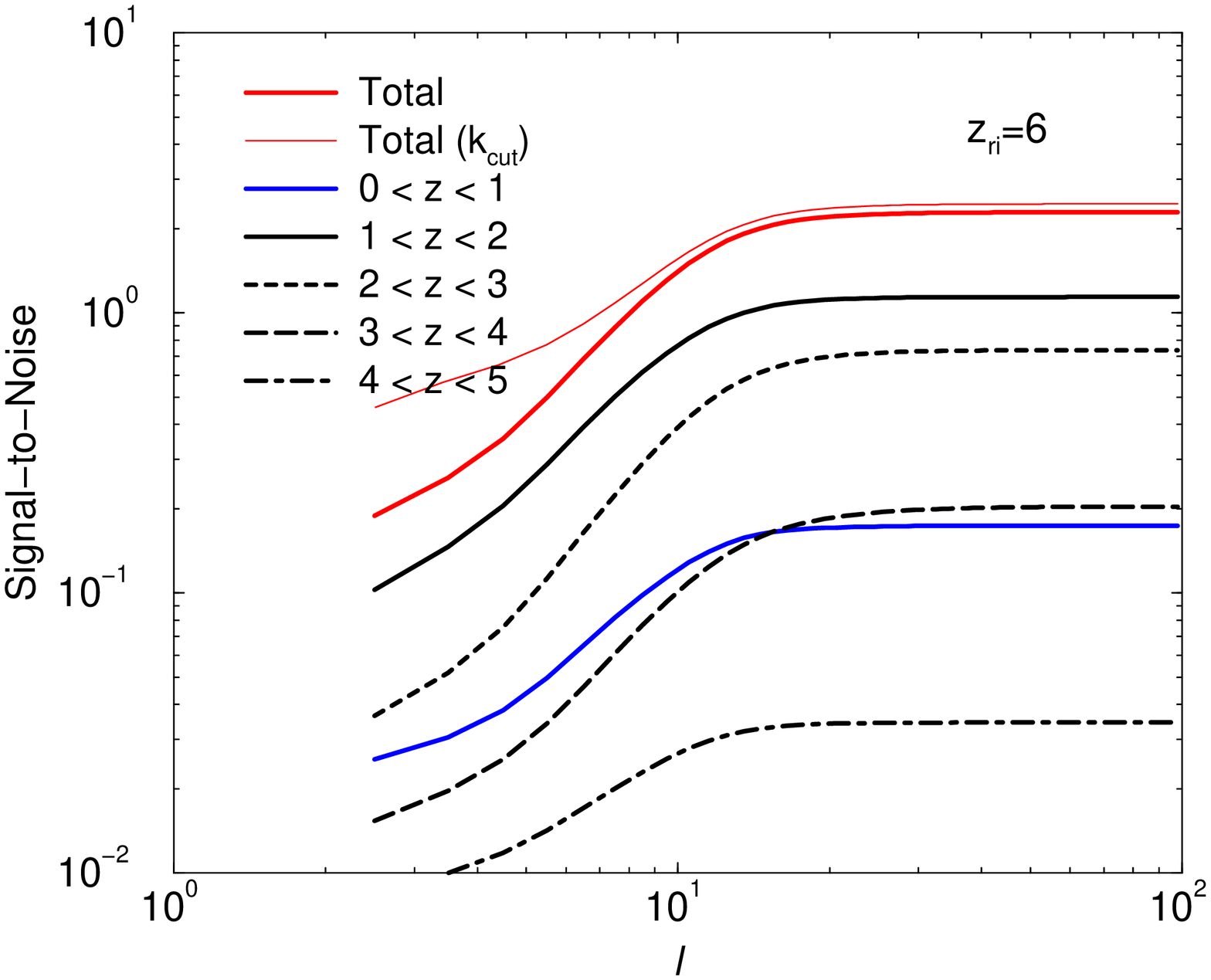,width=3.6in,angle=-90}}
\caption{The cumulative signal-to-noise ratio for a detection of the E-mode polarization-source correlations
as a function of the multipole.
The signal-to-noise ratios peak at a multipole around $\sim$ 10.
The two panels assume a reionization redshift of 4 and 6, while for a reionization redshift 8 and higher
the total signal-to-noise ratio is below unity. The tracer field is divided to redshift bins as labeled on the right panel
assuming tracers, such as galaxies, are uniformly distributed in that bin. We assume a source bias
of unity with respect to the linear density field regardless of the redshift; high redshift sources are
likely to have a bias factor greater than unity (see, for example, Figure~15  of Cooray 2005) and could aid
the detection, but we do not pursue such possibilities here; An increase in the bias factor is likely to be
compensated  by the decreasing density of high-redshift sources. The signal-to-noise ratios plotted here assuming
all-sky maps, but for partial sky coverage, these numbers decrease as $\sqrt{f_{\rm sky}}$, but
coverages less than $\sim$ 50 by 50 degrees are not useful as they do not adequately probe multipoles $\sim$ 10. 
In the right panel, the thin line shows the signal-to-noise ratio with a cut-off in the primordial power spectrum.
A cut-off at a wave number of $5 \times 10^{-4}$ Mpc$^{-1}$ only results in a small increase in the overall signal-to-noise ratio
and a cut-off alone would not significantly boost the ISW-source correlation using polarization maps to a 
level where the detection  is guaranteed.
}
\end{figure*}

For comparison, the angular power spectrum of temperature anisotropies is then given by
\begin{equation}
C_l^\isw = {2 \over \pi} \int k^2 dk P(k)
                \left[I_l^{\rm ISW}(k)\right]^2 \,,
\label{eqn:clexact}
\end{equation}
while the cross-correlation between sources and temperature is simply
\begin{equation}
C_l^{\isw-gal} = {2 \over \pi} \int k^2 dk P(k)
                I_l^{\rm ISW}(k)I_l^{\rm gal}(k) \,,
\label{eqn:clexact2}
\end{equation}
where
\begin{equation}
I_l^{\rm gal}(k)= \int d\rad b(k,\rad) n^{\rm gal}(\rad) j_l(k\rad)
\end{equation}
where $n^{\rm gal}(\rad)$ is the normalized radial distribution of sources (such that $\int dr n^{\rm gal}(\rad)=1$)
and $b(k,\rad) $ is the scale dependent bias factor of these sources relative to the linear density field.
In this calculation, we will assume sources are uniformly distributed in redshift and can be divided to redshift bins
from 0 to 5 at steps of unity. We will assume the bias factor is scale independent and has a value of unity.
Since our calculation is mostly for illustrative purposes to obtain the order of magnitude estimate of the
detectability level of the ISW polarization-galaxy cross-correlation these assumptions are adequate.

The polarization signal is generated by rescattering of the temperature quadrupole. 
We consider the E-mode here since that is induced by scalar fluctuations.
Following Zaldarriaga (1997), the large scale angular power spectrum of the E-mode polarization is
\begin{equation}
C_l^{EE} = {2 \over \pi} \int k^2 dk P(k)
                [I_l^{\rm E}(k)]^2 \,,
\end{equation}
where
\begin{equation}
I_l^{\rm E}(k) = \sqrt{\frac{(l+2)!}{(l-2)!}} \frac{3}{4} \int d\rad g(\rad) Q(k,\rad) \frac{j_l(k\rad)}{(k\rad)^2}
\label{eqn:IE}
\end{equation}
where $g(\rad)=\int d\rad \dot{\tau}e^{-\tau}$ is the visibility function that determines the scattering
probability and the quadrupole that rescatters via electrons in
a reionized Universe has two contributions
\begin{eqnarray}
Q(k,\rad) &=& Q^{\rm SW}(k,\rad)  + Q^{\rm ISW}(k,\rad)  \\
Q^{\rm SW}(k,\rad) &=& -\frac{1}{2} \Omega_m \frac{H_0^2}{k^2} \left(\frac{G}{a}\right)_{\rad_s} j_2[k(\rad_s-\rad)] \nonumber \\
Q^{\rm ISW}(k,\rad) &=& -3 \Omega_m \frac{H_0^2}{k^2} \int_{\rad}^{\rad_s} d\rad' \frac{d}{d\rad'}\left(\frac{G}{a}\right) j_2[k(\rad'-\rad)] \, .
\label{Q}
\end{eqnarray}

In the same manner we described the CMB temperature and source cross-correlation, we can also write the
cross-correlation between a tracer field of the large-scale structure and polarization anisotropies as
\begin{equation}
C_l^{E-gal} = {2 \over \pi} \int k^2 dk P(k)
                I_l^{\rm E-ISW}(k) I_l^{\rm gal}(k) \,,
\end{equation}
where the source related to polarization is now the ISW effect only (such that  in equation~\ref{eqn:IE}, the quadrupole $Q(k,\rad)$ is
replaced by $Q^{\rm ISW}(k,\rad)$ from equation~\ref{Q}).

In Figure~1 (left panel), we show the rms 
quadrupole of SW and ISW effects as a function of redshift. These rms quadrupoles (Hu 1999; Baumann et al. 2003) are calculated through
\begin{equation}
Q_{\rm rms}^2(\rad) = \int \frac{k^2 dk}{2\pi^2} Q^2(k,\rad)  \, .
\end{equation}
For comparison, in Figure~1, we also show the function $g(\rad)$.
The rms quadrupoles, as a function of redshift, show the general behavior of the primordial contribution (SW) and the
relative effect from the ISW effect to CMB polarization. While the ISW contribution decrease with increasing redshift,
the scattering probability increases. On the other hand, rms SW quarupole is essentially the same regardless of redshift;
on the otherhand, due to changes in the size of the last scattering surface an electron sees at high redshift,
$Q^{\rm SW}(k,\rad)$ does change as a function of redshift when considered as a function of the wavenumber (for example,
see Baumann \& Cooray 2003; Seto \&  Pierpaoli 2005).
Thus, if the scattering were to begin at a higher redshift, the overall contribution to polarization would be higher and be dominated by
the SW quadrupole. As long as the reionization redshift is greater than 4, the ISW contribution is no more than 0.01 $\mu$K, while
the SW contribution is at the level of 0.1 $\mu$K or higher depending on the exact redshift of reionization (see, Figure~1 right panel).

For comparison between the temperature and polarization cross-correlation signals, 
it is useful to consider the cross-correlation coefficient $r$ between CMB signals and the tracer field.
 In the case of temperature-galaxy correlation, this coefficient is
\begin{equation}
r_{\rm temp}=\frac{C_l^{\isw-gal}}{\sqrt{C_l^{\rm temp} C_l^{\rm gal}}}
\end{equation}
where $C_l^{\rm temp}$ is total power spectrum of temperature fluctuations and $C_l^{\rm gal}$ is the power spectrum of
sources. Similarly, for polarization, this correlation is
\begin{equation}
r_{\rm pol}=\frac{C_l^{E-gal}}{\sqrt{C_l^{EE} C_l^{\rm gal}}} \, ,
\end{equation}
where now $C_l^{EE}$ is all contributions to polarization both from rescattering at a reionized Universe and from
last scattering surface.

In Figure~2, we summarize the correlation coefficients for polarization-source 
cross-correlation detection for three assumed values of the reionization
redshift:4, 6, 8. While it is clear that reionization happened at a redshift greater than 6, we use $z_{\rm ri}=4$ for illustrative 
purposes here.
In the same figure, we also show the correlation coefficient related to temperature-source detection (bottom right panel).
The correlation coefficient in this case  is at the level of $0.2$ for sources at redshifts less than unity. On the other hand,
the correlation coefficient is smaller ($< 0.1$) for polarization-source correlation when $z_{\rm ri} \geq 6$.
These correlations peak for sources with redshifts between 1 and 2. Sources with redshifts between 2 and 3 have a larger 
correlation relative
to sources between redshifts 0 and 1. Unlike the case with temperature-source correlation coefficients, as a function of the multipole
where one sees a broad range in multipole values for essentially the same coefficient, the correlation coefficient for
polarization-source case decreases rapidly when 
$l > 10$ due to the sharp rise in the polarization power spectrum (see, Figure~1)
associated with primordial polarization
generated by the velocity gradients at the last scattering surface. The correlation coefficient is, in fact, higher
around a multipole of 10 due to the ``dip'' in the EE polarization power spectrum. 

The above comparison indicates that the detection of the polarization-source cross power spectrum
may be challenging, and to quantify this further, we also estimate the
signal-to-noise ratio for its detection 
\begin{equation}
\left(\frac{\rm S}{\rm N}\right)^2 = \sum_l 
f_{\rm sky} (2l+1) \frac{\left(C_l^{E-gal}\right)^2}{C_l^{E-gal} +  \left(C_l^{\rm EE}+C_l^{\rm noise}\right)\left(C_l^{\rm gal}+N_l^{\rm gal}\right)} \, .
\end{equation}
To estimate the maximal signal-to-noise ratio, we ignore noise in CMB polarization map,
$C_l^{\rm noise}$, and shot-noise of  galaxies, $N_l^{\rm gal}$, and concentrate only on the cosmic
variance. This is a safe assumption that at angular scales or multipoles
where this correlation peaks, $l \sim 10$, CMB maps will eventually become cosmic-variance limited
(such as the case now for temperature with WMAP) and, again, shot-noise from sources are not a significant
concern. In Figure~3, we show the cumulative signal-to-noise ratios as a function of the increasing multipole.
As expected the signal-to-noise ratios peak at a multipole around 10 and when all redshift bins are summed, the total
is at the level of $\sim$ 3 ($z_{\rm ri}=4$) and 2  ($z_{\rm ri}=6$) and drops below unity when $z_{\rm ri} > 10$.
These signal-to-noise levels assume all-sky maps of the tracer field and for partial sky coverage, the ratios decrease as $\sqrt{f_{\rm sky}}$.
Even for all-sky maps, the detection of the cross polarization-source signal is marginal. 

This signal-to-noise ratio, however, is not significantly less than unity. This suggests that, while marginal, other reasons
may determine if the correlation can be detected or not. For example, while we have not explored the large range of models
suggested for dark energy, there may be a favorable dark energy model already in the literature or
reasonably conceived to ``boost'' the ISW signal at redshifts between 1 and 3. Similarly, $C_l^{\rm EE}$ may be smaller
at small multipoles if some suggestions related to the apparent lack of power in CMB temperature maps were to hold,
such as those involving a cut-off. In Figure~1, we show an example case where we consider the primordial power
to be zero at wavenumbers $k_{\rm cut} < 5 \times 10^{-4}$ Mpc$^{-1}$. Such a cut-off explains the apparent
low power of the quadrupole and the octupole of the temperature anisotropy maps (e.g., Kesden et al. 2003). 
While a better characterization of the cut-off, if it exists,
 may be interesting to understand physics during the early Universe (e.g., Hannestad \& Mersini-Houghton 2005),
techniques to further study it are limited (Kesden et al. 2003).
The ISW-polarization cross-correlation may provide additional information, though we find that the overall
increase in the signal-to-noise ratio with a cut-off is not significant.
Finally, while we have considered the angular cross power spectrum,
there may be more suitable techniques to look for the ISW polarization-source correlations.
Such techniques could involve looking for correlations directly in the pixel space, techniques to account for
primordial signal in the polarization map based on additional correlations, among others. 

Any attempts to detect this correlation also require wide-field maps of the large-scale structure
at redshifts between 1 and 3. While such information already exists at low redshifts, with surveys such
as the Sloan Digital Sky Survey, high redshift maps will soon become available with
planned sky surveys with instruments such as the Large Synoptic Survey Telescope (LSST; Tyson et al. 2002).
Studies on how large-scale structure involving billion or more galaxies  from LSST
can be applied for astrophysical and cosmological studies are already underway (e.g., Zhan et al. 2005).
Improved polarization maps, from WMAP data and the upcoming Planck mission, can easily be combined with 
imaging data from LSST, especially on large-scale structure around redshifts 1 to 3, 
to look for the suggested cross-correlation. In addition to galaxies, sources that span the required redshift ranges include
quasars and AGNs at other wavelengths. While the detection requires favorable conditions even
a non detection may be useful to constrain properties of dark energy at such high redshifts.
This is also useful again in the context that most other cosmological probes of
dark energy, such as galaxy cluster number counts, consider redshifts below unity
or out to at most 1.7 (such as Type Ia supernovae).  

Beyond dark energy, this test has the most power
in constraining the overall redshift of reionization. For example, a detectable signal would generally require that the
Universe reionized at a late time, such as through quasars at a redshift $\sim 6$. On the other hand current
determinations  of the optical depth argue for a much higher redshift for reionization with optical depth
$\sim$ 0.17. The latter has large uncertainties both in the measurements as well as in the interpretation through
degeneracies between other parameters. With high signal-to-noise E-mode polarization maps cross-correlation studies
can be considered and if a detection  were to be found, this should generally argue for a low reionization redshift.

\section{Conclusions}

The cross-correlation between cosmic microwave background (CMB) temperature anisotropies and the
large scale structure (LSS) traced by the galaxy distribution, or sources at different wavelengths,
is now well known. This correlation results from the integrated Sachs-Wolfe (ISW) effect in CMB anisotropies
and provides a measure of dark energy and its physical properties. In a reionized universe,
electron scattering at low redshifts leads to a large-scale polarization contribution.
In addition to the primordial quadrupole, involving anisotropies at the last scattering surface,
the ISW quadrupole rescatters and contributes to large-scale polarization signal.
Thus, in principle, the large-scale polarization bump in the E-mode 
should also be correlated with the galaxy distribution.
Unlike CMB-LSS correlation that peaks for tracers at low redshifts,
due to the decreasing visibility function at low redshifts and the decreasing ISW contribution at high redshifts,
the correlation peaks mostly at redshifts between 1 and 3. 
Under certain conditions, mostly involving a low optical depth to reionization, the
polarization signal should be detectable, though signal-to-noise ratios involving an all-sky
map of the galaxy distribution is not expected to be high and is likely to be at the level of two to three
if the Universe reionized at a redshift around 6. If the Universe reionized at a redshift higher than 8,
it is unlikely that this correlation will be detectable. Unknown physics related to the dark energy
as well as still uncertain issues related to the large angular scale CMB and polarization anisotropies, however,
may compel attempts to measure this correlation using upcoming CMB polarization E-mode maps.

\section{Acknowledgments}
We thank Levon Pogosian, Pier-Stefano Corasaniti, and Niyesh Afshordi for useful comments on an early draft of this paper.
AM is supported by MURST through COFIN contract no. 2004027755.

\end{document}